\documentclass{ws-mpla}
\usepackage[super]{cite}
\usepackage{graphicx}
\usepackage[utf8]{inputenc}
\usepackage{amsmath}
\usepackage{amsfonts}
\usepackage{amssymb}
\usepackage{hyperref}
\begin{document}

\markboth{Samuel Kov\'a\v{c}ik}
{$R^3_\lambda$ inspired black holes}
%%%%%%%%%%%%%%%%%%%%% Publisher's Area please ignore %%%%%%%%%%%%%%
\catchline{}{}{}{}{}
%%%%%%%%%%%%%%%%%%%%%%%%%%%%%%%%%%%%%%%%%%%%%%%%%%%%%%%%%%%%%%%%%%%
\title{$R^3_\lambda$ inspired black holes}

\author{\footnotesize Samuel Kov\'a\v{c}ik}

\address{School of Theoretical Physics,\\ Dublin Institute for Advanced Studies, \\10 Burlington Road, Dublin 4, Ireland.\\
skovacik@stp.dias.ie}

\address{Faculty of Mathematics, Physics and Informatics,\\
Comenius University Bratislava,\\
 Mlynsk\'a dolina, Bratislava, 842 48, Slovakia.}

\maketitle

\pub{Received \today}{%Revised \today
}

\begin{abstract}
We study a black hole with a blurred mass density instead of a singular one, which is caused by the noncommutativity of 3-space. Depending on its mass, such object has either none, one or two event horizons. It possesses properties, which become important on a microscopic scale, in particular, the Hawking temperature does not increase indefinitely as the mass goes to zero, but vanishes instead. Such frozen and extremely dense pieces of matter are good dark matter candidates.

\keywords{Noncommutative quantum mechanics; microscopic black holes; dark matter.}
\end{abstract}

\ccode{PACS Nos.: 04.60.Bc, 02.40.Gh}
\vspace*{.25cm}
\small{\ Ref. number: DIAS-STP-16-09}

\section{Introduction}

Noncommutative (NC) theories are built on spaces whose coordinates do not commute and therefore one cannot localize their points with an arbitrary precision. This is believed to be an artifact of quantum gravity and results into a plenty of novel properties, for example a natural UV cutoff. Often are of interest NC spaces with a different dimensionality and geometry than that of $\textbf{R}^3$ space, for example the Moyal plane, the fuzzy sphere or the fuzzy torus, see \cite{NC1,NC2, NC3} and references therein.  

Construction of a NC Euclidean 3-space $\textbf{R}^3_{\lambda}$ and quantum mechanics (QM) on it has been postulated in \cite{Jabbari} and developed in more detail in \cite{ncqmLRL, ncqmHatom1,ncqmHatom2,ncqmVelo, mm} {}. A notable feature of the model is that it often allows exact results, many of them being very close to those known in ordinary QM. 

Among the objects sensible to introducing a short scale structure described by $\lambda$ are microscopic black holes with radius of that order $r_h \sim \lambda$. In the classical theory, as they evaporate by Hawking radiation \cite{Hawking} their radius eventually becomes infinitely small and their temperature becomes infinitely high. It is interesting to study how the UV cutoff of $\textbf{R}^3_\lambda$ manifest itself in this scenario. The appropriate theory to investigate this would be the one of quantum gravity, which is yet to be found, or at least properly understood. Therefore we settle for a semiclassical description and implement some of the results from NC QM into the classical theory of gravity. 

This method was developed by Nicolini, see \cite{Nicolini} for a review, who dubbed it 'NC inspired'. More details on NC inspired cosmology and gravity could be found \cite{infl1, ncmbh1,ncmbh2,ncmbh3,ncmbh4,ncmbh5,ncmbh6}{}, the concept of generalized uncertainty principle in a similar context has been analyzed as well, see \cite{Arraut1, Arraut2, rev9}{}. A remarkable series of papers by Dymnikova et al. deals with a subject of regular black holes, see \cite{rev3,rev4,rev7,rev8, rev11, rev12, rev13, rev14}{}.

This paper is organized as follows. In Section \ref{sec:NC} we derive a NC analogue of the Dirac delta function in $\textbf{R}^3_{\lambda}$, complete it into stress-energy tensor $T^\mu _\nu$ and solve the Einstein field equations (EFE) for it. This solution is analyzed, focusing mostly on the event horizon(s) and the Hawking temperature, in Section \ref{sec:ehhr}. Section \ref{sec:PC} contains physical consequences of the model and final conclusions.

\section{Noncommutative space, (blurred) delta function and EFE}
\label{sec:NC}
We will study a model of 3 dimensional rotational invariant NC space described by 
\begin{equation} \label{NCrel}
[\hat{x}_i, \hat{x}_j] = 2 i \lambda \hat{x}_k \varepsilon^{ijk}\, ,
\end{equation}
where $\varepsilon^{ijk}$ is the Levi-Civita symbol and $\lambda$ is a constant with the dimension of length, describing the scale of noncommutativity. It is not fixed within our model, but could be expected to be (approximately) the Planck length.

There are several ways how to satisfy \eqref{NCrel}, for example see \cite{Berezin,Bosonic,Grosse,ncqmHatom1,Madore,Presnajder}{}. We will employ the bosonic operator approach which was previously used in \cite{ncqmLRL,ncqmHatom1,ncqmHatom2,ncqmVelo, mm} and is well suited for 3 dimensional rotational invariant problems.

Let us define two sets of bosonic creation and annihilation operators satisfying 
\begin{equation} \label{bos}
[\hat{a}_\alpha , \hat{a}^+_\beta] = \delta_{\alpha \beta}\, ; \, \alpha,\beta =1,2 
\end{equation}
and acting in an auxiliary Fock space $\mathcal{F}$ spanned on normalized states $| n_1, n_2 \rangle = \frac{(\hat{a}_1^+)^{n_1}(\hat{a}_2^+)^{n_2}}{\sqrt{n_1!n_2!}}| 0,0 \rangle$, where $| 0,0\rangle = |0\rangle$ is the vacuum state annihilated by both $\hat{a}_\alpha$. It is convenient to define their dimensional versions as $\hat{z}_\alpha = \sqrt{\lambda} \hat{a}_\alpha, \, \hat{z}_\alpha^+ = \sqrt{\lambda}\hat{a}^+_\alpha$. Using these (and the Pauli matrices $\sigma^i$), we can define the (Cartesian) coordinates satisfying \eqref{NCrel} and the radial coordinate \footnote{Their relation is $\hat{r}^2 -\hat{x}^2 = \lambda^2$, as can be easily checked.} as
\begin{equation} \label{NCx}
\hat{x}_i =  \sigma^i_{\alpha \beta} \hat{z}^+_\alpha\hat{z}_\beta , \ \hat{r}= \hat{z}^+_\alpha \hat{z}_\alpha +\lambda \ .
\end{equation}

Fock space states $| n_1, n_2 \rangle$ are $\hat{r}$ eigenstates with eigenvalues of $\lambda (n_1+n_2 +1)$, the vacuum state $|0,0\rangle \equiv |0\rangle$ is the state with the minimal eigenvalue, it corresponds to the origin of the coordinate system.

This construction can be understood either as a sequence of concentric fuzzy spheres, or as a Hopf fibration $S^3 \rightarrow S^2$ from quantized $C^2_\lambda$ to $R^3_\lambda$, see \cite{mm}{}.

Coherent states play an important role in the ordinary quantum mechanics and are crucial in NC theories as well, see \cite{cs1,cs2,cs3,cs4,cs5}{}. A coherent state is well localized wave packet which minimizes the uncertainty relation and is defined as an annihilation operator eigenstate $\hat{a} |\alpha\rangle = \alpha |\alpha\rangle$. Such states can be constructed as $|\alpha\rangle = e^{- \frac{|\alpha|^2}{2}} e^{\alpha \hat{a}^+} |0\rangle $ and used as an overcomplete set of states in $\mathcal{F}$,  \cite{Perolomov}{}. Overlap of two such states is $\langle \alpha|\beta \rangle = e^{ -\frac{|\alpha|^2 + |\beta|^2}{2} +\bar{\alpha}\beta}$. We are interested in a state well localized at the origin, which follows from 
\begin{equation} \label{rho0}
\tilde{\rho}(z) = | \langle z|0 \rangle |^2 =  e^{-\frac{|z|^2}{\lambda}} = e^{-\frac{r-\lambda}{\lambda}}\, .
\end{equation}
 
Let us pause for a moment to make a few remarks here. First of all, we define taking $\lambda \rightarrow 0$ as the commutative limit (RHS of \eqref{NCrel} vanishes, as in the ordinary space). It is easy to see that in this limit the RHS of \eqref{rho0} vanishes everywhere but at the point $r=0$, it behaves like the Dirac delta distribution with the source located at the coordinate origin. It is therefore natural to call $\tilde{\rho} \propto e^{-\frac{r}{\lambda}}$ a NC delta distribution or a blurred delta distribution (located at the coordinate origin).

Note that $\tilde{\rho}$ in \eqref{rho0} is dimensionless, a dimensional one will be denoted $\rho$. Since the rest of calculations will be done using the ordinary (not NC) calculus, we will normalize $\rho$ with respect to ordinary integration instead of trace norm, yielding
\begin{equation} \label{rho}
 \rho (r)= \frac{M}{8\pi\lambda^3}e^{-\frac{r}{\lambda}} \, .
\end{equation}
 
In the paper by P. Nicolini \cite{Nicolini}{}, which served as a main inspiration for ours, a similar line of reasoning was used. However, the starting point was a two dimensional NC space and the resulting density was generalized into three dimensional only afterwards, yielding $\rho \propto  e^{ - r^2/ \lambda_{2D}^2}$. As we have shown, a direct three dimensional derivation based on \eqref{NCrel} gives a different proportionality \eqref{rho}.

Models that do not postulate a NC structure of the underlying space can also lead to a "blurred delta" distribution. For example in the aforementioned papers on regular black holes it is assumed that $\rho \propto  e^{ - r^3 / r_0^3}$, which is being motivated by the vacuum polarization in a gravitational field.

To see how is the UV regularization of $\textbf{R}^3_\lambda$ realized in the context of Hawking radiation of microscopic black holes we will take \eqref{rho} as the matter density and complete it into stress-energy tensor. We focus on uncharged nonrotating black holes, so we expect the solution to become Schwarzschild-like in the $\lambda \rightarrow 0$ limit. This encourages us to use a diagonal ansatz for the metric tensor with $g_{tt}=-g_{rr}^{-1}$, therefore we seek only a single function $f(r)$ such that $g_{\mu \nu} = \mbox{diag} \left( f, -f^{-1}, r^2, r^2 \sin^2 \theta \right)$ (with coordinates  $(t,r,\theta,\varphi)$ and signature $(-,+,+,+)$. We often set unimportant constants equal to $1$ and omit writing arguments).

For the same reason we are expecting a diagonal $T^\mu_\nu$ with $T^t_t = - \rho$ and $T^r_r=T^t_t$ (which follows from the EFE). Because of the chosen ansatz, $T^r_r=T^t_t$ is fixed as well (this also can be seen from the EFE). The other two components follow from the conservation law $T^{\mu \nu}{}_{;\nu}=0$. For $\mu=\theta$ we get $ T^\theta_\theta= T^\varphi_\varphi =: p_\perp$, for $\mu=r$ we get $p_\perp = - \frac{r}{2}(\partial _r \rho + \frac{2}{r}\rho)= -\rho - \frac{r}{2} \partial _r \rho$. Adding this we obtain $ T^\mu_\nu= \left( -\rho, p_r, p_\perp, p_\perp \right)$, where $p_r = -\rho$ and $p_\perp$ is defined above.

This is in agreement with the Schwarzschild-like class of solutions that has been studied in \cite{rev4, rev5, rev6}{}. The generating action has been discussed in \cite{rev2}{}.  

Since we are looking for a single function $f(r)$ we only need one of the EFE, let us take $G_t^t = 8 \pi T_t^t$. From it the solution follows as
\begin{equation} \label{eqForF}
\frac{ 1 + f(r) + r f'(r)}{r^2}=\frac{M}{\lambda^3}e^{-\frac{r}{\lambda}} \rightarrow f(r) = -1 - e^{-\frac{r}{\lambda}}\frac{M}{r}\left( \frac{r^2}{\lambda^2}+\frac{2r}{\lambda}+2 \right)+\frac{C}{r} \, .
\end{equation}

Recall that $g_{tt}=f$, therefore if we want the solution to approach Schwarzschild solution for $r \gg \lambda$, we need to set $C=2M$. For the rest of this paper we will be needing only the time component of the metric tensor,
\begin{equation}\label{G}
g_{tt}(r;\lambda,M) =  -1 + \frac{2M}{r} - e^{-\frac{r}{\lambda}}\frac{M}{r}\left( \frac{r^2}{\lambda^2}+\frac{2r}{\lambda}+2 \right) \, .
\end{equation}
 
Let us pause for a brief comment. The stress-energy tensor violates the weak energy condition (see \cite{Carroll}) for $r<2\lambda$, which signalizes a quantum repulsion (preventing the matter from collapsing into a singularity \cite{rev4, rev6}{}). The strong energy condition is violated between black hole horizons. This might either be an artifact of the semiclassical approach, or a signal of negative energies present. In \cite{Barcelo} it is argued that the energy conditions are becoming obsolete, being violated even in classical theory; namely, local conditions are not satisfied for scalar fields with nonminimal coupling to gravity, and even averaged conditions do not always hold if the field reaches trans-Planckian values. This issue has already been addressed in a more general context in \cite{rev4, rev5, rev6}{}.

The particular form of the stress-energy tensor followed from the postulate of $T^t_t$ and from the requirement of Schwarzschild-like form of $g_{\mu \nu}$. However, since we will be using only the $g_{tt}$ component our results hold also for different completions of $T_{\mu \nu}$, which might be free of (strong condition) violations. Generality with respect to a different choice of (regular) matter distribution has been discussed in \cite{rev7}{}, we will address this issue in Section 4.

\section{Event horizon(s) and Hawking radiation}
\label{sec:ehhr}
Before we begin this section, let us summarize the expectations for black holes in $R_\lambda^3$. Theories in this space have two (dual) properties – both the notion of a point and infinite energies are absent. The black holes should be unable to shrink to a point and their temperature should be prevented from reaching infinite values. Such properties have also been observed in general studies of regular black holes \cite{rev3,rev4,rev7,rev8,rev11,rev12,rev13,rev14, rev5,rev10} or NC inspired black holes with a different choice of $\rho$ \cite{Nicolini}{}.

Event horizons are solutions of the equation $g_{tt}(r) = 0$. For an ordinary Schwarzschild black hole there is only a single solution $r=2M$, now there are two, one or zero solutions, depending on the value of $M$ (see Figure 1). In the case there are two, let us denote them $r_-, r_+$ ($r_- < r_+$). Existence of multiple horizons is a general property, see for example the aforementioned references on regular black holes.

\begin{figure}[b]\label{figure1}
 \includegraphics[width=1\textwidth]{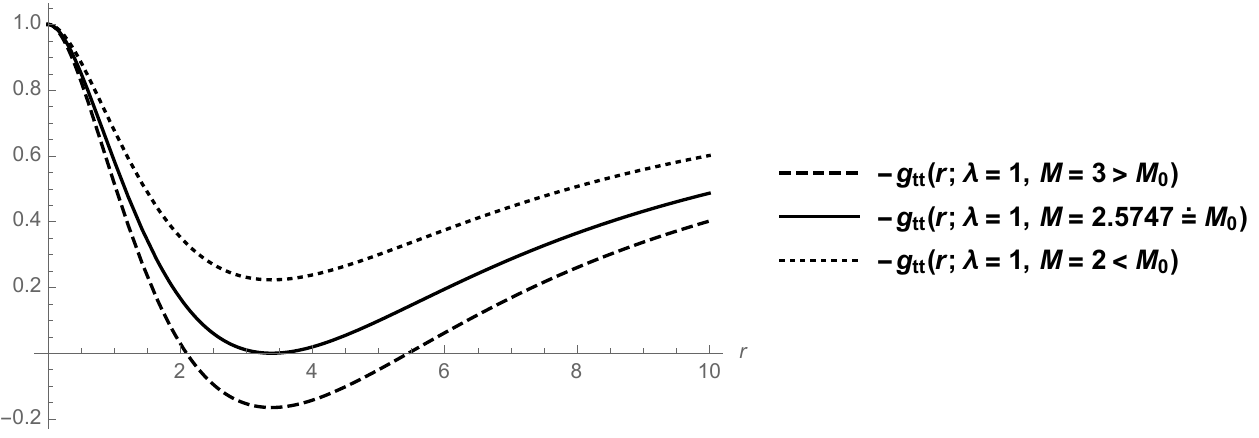}
 \caption{$-g_{tt}(r)$ for  $\lambda=1$ and different values of $M$.}
\end{figure}
When the mass is large ($M\gg\lambda$), there are two horizons, one near the singularity ($r_- \approx 0$) and the other near the classical horizon ($r_+ \approx 2M$). As $M$ gets smaller, these two surfaces move towards each other and meet for $M=:M_0$ at $r=:r_0$. A black hole with the mass $M_0$ and a single horizon at $r_0$ is minimal, since for any smaller $M$ there is no horizon at all, minimal black hole is the smallest (and lightest) possible black hole. The values of $M_0,r_0$ can be obtained numerically
\begin{equation}\label{Mr}
M_0 \stackrel{.}{=}  2.57 \lambda, \, r_0 \stackrel{.}{=}  3.38 \lambda\,.
\end{equation}
The Hawking temperature of a minimal black hole is zero. This follows from the fact that it is proportional to the surface gravity at the (outer) horizon $\kappa = - \frac{g_{tt}'(r_0)}{2}$, see \cite{Hawking}{}, which has to be zero, since $-g_{tt}(r)$ reaches its minimum there. The black hole becomes frozen and evaporation ceases when the minimal mass $M=M_0$ is reached.

%RESTORE THIS
\begin{figure}[!b] \label{fig2}
 \includegraphics[width=1\textwidth]{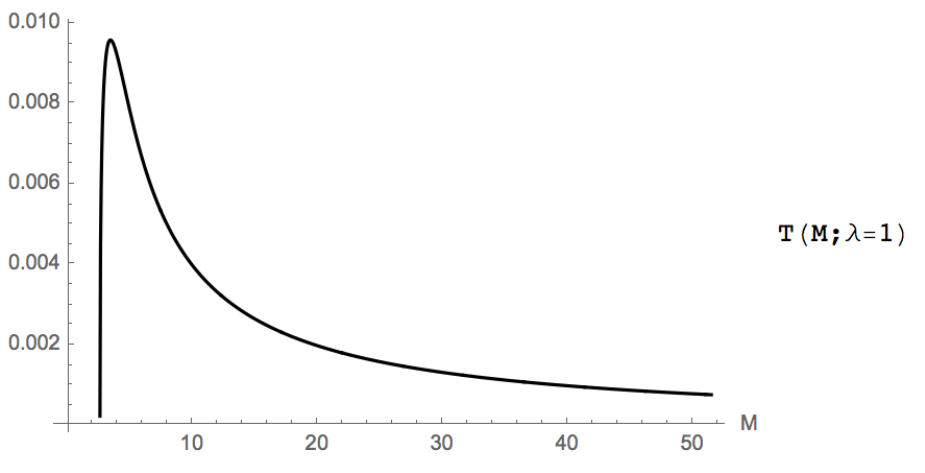}
 \caption{The Hawking temperature as a function of black hole's mass.}
\end{figure}

Note that infinite temperatures are avoided (Figure 2). From a dimensional analysis we can see that the maximal reached temperature (denoted $T_m$) is proportional to $\lambda^{-1}$.  To find the constant of proportionality, let us first factorize out the mass from $g_{tt}$
\begin{equation} \label{Tg}
g_{tt}(r;\lambda,M) = -1 + M \tilde{g}(r;\lambda) \, ,
\end{equation}
where $\tilde{g}(r)$ does not depend on $M$. At the (outer) horizon $\tilde{g}(r_+) = M^{-1}$, and 
\begin{equation} \label{tg}
g'_{tt}(r_+)= M \tilde{g}'(r_+) = \frac{\tilde{g}'(r_+)}{\tilde{g}(r_+)}\, .
\end{equation}
This is, up to a multiplicative constant, equal to the Hawking temperature. To find $r_+$, for which this achieves extremum we need to solve $\partial_{r_+} g'_{tt}(r_+) =0$. This can be, again, done numerically (choosing $\lambda=1$), finding that the extremum is reached as $g'_{tt}(r_+ \stackrel{.}{=}6.54) \stackrel{.}{=} - 0.12$. Plugging this into the relation for the temperature we obtain
\begin{equation} \label{Tm}
T_{m} \stackrel{.}{=} \frac{\hbar c}{4 \pi k_B}\frac{0.12}{\lambda} , \, \frac{\hbar c}{4 \pi k_B}\stackrel{.}{=}0.18 \times 10^{-3}m K.
\end{equation}
It can be observed in Figure 2 that the temperature grows very rapidly for $M \gtrsim M_0$. It is therefore interesting to investigate what happens after adding a small mass $\delta M \ll M_0$ into a minimal black hole.

To answer this question we use the decomposition \eqref{Tg}. Let us denote the increment in radius $\delta r$, horizon condition after adding a small mass $\delta M$ reads
\begin{equation}\label{hor1}
 -1 + (M_0 + \delta M) \tilde{g}(r_0 + \delta r;\lambda) = 0.
\end{equation}
Truncating the Taylor expansion of \eqref{hor1} we obtain
\begin{equation}
\tilde{g}(r_0 + \delta r;\lambda) \stackrel{.}{=} \underbrace{\tilde{g}(r_0;\lambda)}_{M_0^{-1}} + \delta r \underbrace{\partial_r \tilde{g}(r_0;\lambda) }_0 + \frac{1}{2}\delta r ^2\partial_r^2 \tilde{g}(r_0;\lambda) \, .
\end{equation}
Inserting this back into \eqref{hor1} we arrive to
\begin{equation} \label{Dr}
\delta r \stackrel{.}{=} \pm \sqrt{\frac{-2 \delta M }{M_0^2\partial_r^2 \tilde{g}(r_0;\lambda)  }} \, .
\end{equation}
Evaluating for $M_0$ and $r_0$ as given in \eqref{Mr} yields $\delta r \stackrel{.}{=} \pm 2.54 \sqrt{ \lambda \delta M}$ (there are two symmetric solutions because we have truncated the Taylor expansion after the quadratic term).

We can now determine the temperature of the resulting black hole 
\begin{equation}
T(r_0 + \delta r) \stackrel{.}{=} \overbrace{T (r_0)}^0 + \partial_r T(r_0) \delta r 
\end{equation}
\begin{equation*}
= - \frac{M_0 \tilde{g}''(r_0)}{4 \pi} \delta r   \stackrel{.}{=} \frac{1}{4 \pi}\frac{1}{M_0} \frac{2\delta M}{\delta r} \stackrel{.}{=}\frac{1}{2\pi} \frac{\sqrt{\delta M}}{ \ 6.53\ \lambda^{3/2} } \, ,
\end{equation*}
 Recovering constants for a moment
\begin{equation}
T(M_0 + \delta M) \stackrel{.}{=} \frac{\sqrt{\delta M}}{41.01\ \lambda^{3/2} }\frac{\hbar c}{k_B} \, .
\end{equation}
It is useful to express this with respect to the maximal temperature
\begin{equation} \label{THTM}
\frac{T(M_0 + \delta M)}{T_m} \stackrel{.}{=}2.55 \sqrt{\frac{\delta M}{\lambda}}\stackrel{.}{=}4.09 \sqrt{\frac{\delta M}{M_0}} \, .
\end{equation}
We can see that for $\delta M \ll M_0$ the black hole does not reach its maximum temperature, only a small fraction of it (which is, absolutely speaking, still immense). 

%The last question of this section will be whether does the temperature reach the maximum value if we merge two minimal black holes together? When we have $M=2M_0$ we are in a region where we can safely take $r_+ = 2M = 4M_0 \stackrel{.}{=} 10.28 \lambda$. This is larger than the value $r_+ = 6.54 \lambda$ for which the temperature reaches maximum, therefore the maximum will be reached, when the new black hole evaporates from the radius of $10.28 \lambda$ to $6.54 \lambda$. 

\section{Physical consequences and conclusion}
\label{sec:PC}

To be able to evaluate physical consequences let us assume that $\lambda \sim l_{\mbox{Planck}} \stackrel{.}{=} 1.62 \times 10^{-35}m$, as is usually done for NC spaces (scaling rules for a different choice will be included). Most sensitive to introducing the noncommutativity are microscopic black holes, with a radius of the order of a few $\lambda$. The most important case is the minimal black hole, let us denote it mBH.

According to \eqref{Mr} a mBH should have radius  $r_0 \stackrel{.}{=} 5.48 \times 10^{-35}m$ (we can take the cross section to be $\sigma=\pi r_0^2 \stackrel{.}{=} 9.43 \times 10^{-69}m^2$) and mass $M_0 \stackrel{.}{=} 5.59 \times 10^{-8}kg$ ($r_0,M_0$ scale as $\lambda$). Furthermore the maximal temperature $T_{\mbox{m}}$ is $1.33 \times 10^{30}K$, which is two orders below the Planck temperature (this scales as $\lambda^{-1}$).

Considering these numbers, mBH (or microscopic black holes in general) are possible cold dark matter constituents (this idea appeared 30 years ago \cite{mBHDM}{}, but is difficult to test). They are cold and dark (since their radiation froze out), have extremely small cross section and are heavy enough so only a small concentration $n_{mBH} \stackrel{.}{=}4.25\times 10^{-20} m ^{-3}$ is needed to make up for the observed dark matter mass density $\rho_{DM} \stackrel{.}{=} 2.38 \times 10^{-27} kg \, m^{-3}$  (this scales as $\lambda^{-1}$), see \cite{DMexp}. Dark matter density is uniform only on cosmological scales, there is more of it in galaxies (by factor $10^5-10^6$, see \cite{Weber}{}, possibly even more within solar systems). The idea of mBH as dark matter candidates and their formation has been discussed already, for example in \cite{rev2, rev3, rev12}{}. 

The cross section of mBH is small enough for them not to interact with each other, however it is still possible for them to be hit by another particle. Let us assume that a mBH gets hit by a proton and absorbs it, what would happen? Since the mass of the proton is significantly smaller than $M_0$ we can use eq. \eqref{THTM}, for this example $\frac{\delta M}{M_0} \stackrel{.}{=} 2.98 \times 10^{-20}$. The resulting microscopic black hole will warm up to $7.06 \times 10^{-10}$ of $T_m$, which is $9.39 \times 10^{20}K$ (this scales as $\lambda^{-\frac{3}{2}}$), two orders below the energy of ultra-high-energy cosmic rays which are being observed, see \cite{rays}{}. Had the $\lambda$ been shorter than the Planck length, a radiation of a microscopic black hole after consuming a proton could account for such rays. It should be noted here that it might be more correct to consider mBH-electron or mBH-quark collision instead, since the proton is significantly larger than mBH. Ultra-high-energy bursts has been considered, in the context of graviatoms, in \cite{rev14}{}. 

It is important to note that in the considered case the energy of radiation exceeds the energy of the consumed particle. The possible scenario is that the energy will be radiated in one or two quanta and the resulting object will end with $M<M_0$, it will have no horizons and stops being a black hole \footnote{Gravitational solitons without horizons have been called G-lumps in \cite{rev3, rev7, rev11,rev12}{}.}. Then it will be moving through the space as an extremely dense lump of matter and collect additional mass until it reaches the mass $M_0$ and becomes mBH again. 

Such object, let us name it \textit{gravimond}, lives in cycles: first it is a mBH with mass $M_0$. Then, after it absorbs a particle its radiation is reignited as $M>M_0$. Shortly after it stops being a black hole, since so much energy has been radiated that $M<M_0$, it becomes an extremely dense object (almost a black hole), which needs to capture additional mass to become mBH again. The period of these cycles is unknown and largely depends on the location of such object (how often does it get to interact with other matter). 

\textbf{Conclusion:} The paper analyzed (microscopic) black holes with a blurred mass density, instead of a singular one. Such matter density originated from considering a NC structure of 3-space, yet the following calculations have been done using the ordinary calculus and the general theory relativity. There are many cases in the history of physics advocating for a semiclassical approach, just recall the Bohr's derivation of the Rydberg's formula. We do not expect our results to be as exact, but merely to give a hint of what to expect from a proper quantum theory of gravity. Since some of the features persist also in full NC approach, for example existence of the minimal possible event horizon radius can be compared to the minimal area $A\approx 4\pi l_{\mbox{Planck}}^2$ of the event horizon modeled as a fuzzy sphere in \cite{Dolan}{}, it is plausible that other features will hold in a full NC approach as well. 

%Our results are in a general agreement with those of similar models either originating from different NC spaces or using the context of de Sitter-Schwarzschild (regular) black holes, loop quantum gravity, generalized uncertainty principle \cite{Nicolini, rev2,rev3, rev4,rev5, rev6, rev7, rev8, rev9,rev10, rev11,rev12,rev13,rev14,rev15} and references therein. Our effort was to extend the results of \cite{ncqmLRL, ncqmHatom1,ncqmHatom2, ncqmVelo, mm} and to see how is the UV regularization, which took a very explicit form in these works, realized in the context of Hawking radiation of microscopic black holes.

Our results (for example multiple horizons or vanishing temperature) are in a general agreement with those of similar models either originating from different NC spaces \cite{Nicolini, Dolan, ncmbh1,ncmbh2,ncmbh3,ncmbh4,ncmbh5, ncmbh6,rev2} or using the context of de Sitter-Schwarzschild (regular) black holes \cite{rev3,rev4,rev7,rev8,rev11,rev12,rev13,rev14, rev5,rev10}, generalized uncertainty principle \cite{rev9, Arraut1,Arraut2}, loop quantum gravity \cite{rev6}  and references therein. Our effort was to extend the results of \cite{ncqmLRL, ncqmHatom1,ncqmHatom2, ncqmVelo, mm} and to see how is the UV regularization, which took a very explicit form in these works, realized in the context of Hawking radiation of microscopic black holes.

A novel point of our work is that our starting point was $R^3_\lambda$, a NC space as close to ordinary $R^3$ as possible. We have shown that this assumption leads to a different mass density \eqref{rho} as when generalizing from a lower dimensional NC space. The resulting black hole behavior coincide with that of other regular black holes. This generality has been discussed in \cite{rev7}{}, but it can be understood from the context of this paper as well. 

It is possible to imagine that a different model of NC Euclidean 3-space would provide a different matter density than \eqref{rho}. Following the same steps as in this paper, it will come down to solving a differential equation $r f' = R-F$ where $R=r^2 \rho$ and $F=1+f$. Assuming regularity of the mass density and finiteness of the mass we have  $R(0) =F(0) = R(\infty)=0$. Requiring Schwarzschild solution far from the origin (but a regular one close to it) gives $F(r \gg 0) \propto r^{-1}$, which decays slower than $R\propto r^a$ with $a<1$, therefore far from the origin is $F>R$. On the other hand, from the same differential equation follows that close to the origin is $R>F$. Then there has to be at least one point $r=r_M$ for which $R(r_M) = F(r_M)$ and where $f'(r_M)=0$. Therefore the solution $f$ will always have a similar profile as in Figure 1, with the mass of the source determining the number of the roots and therefore also the number of the horizons (even more than two for more complicated non-monotonic functions $\rho$).

\subsection*{Acknowledgment}

I would like to thank P. Pre\v{s}najder (my PhD supervisor) and V. Balek for their valuable comments and corrections and also the reviewer for pointing out a list of interesting references related to this work. This research was partially supported by COST Action MP1405 and UK/92/2015.

\bibliographystyle{ws-mpla}

\end{document}